\documentclass[twocolumn,twoside,slac]{revtex4}
\usepackage{graphicx}
\usepackage{fancyhdr}
\begin{document}

\title{Transparent Persistence with Java Data Objects}

\author{Julius H\v{r}ivn\'{a}\v{c}}
\affiliation{LAL, Orsay, France}

\begin{abstract}
Flexible and performant Persistency Service is a necessary component of any HEP
Software Framework. The building of a modular, non-intrusive and performant 
persistency component have been shown to be very difficult task. In the past, 
it was very often necessary to sacrifice modularity to achieve acceptable 
performance. This resulted in the strong dependency of the overall Frameworks 
on their Persistency subsystems.

Recent development in software technology has made possible to build a 
Persistency Service which can be transparently used from other Frameworks. 
Such Service doesn't force a strong architectural constraints on the overall 
Framework Architecture, while satisfying high performance requirements. Java 
Data Object standard (JDO) has been already implemented for almost all major 
databases. It provides truly transparent persistency for any Java object 
(both internal and external). Objects in other languages can be handled 
via transparent proxies. Being only a thin layer on top of a used database, 
JDO doesn't introduce any significant performance degradation. Also 
Aspect-Oriented Programming (AOP) makes possible to treat persistency as 
an orthogonal Aspect of the Application Framework, without polluting it 
with persistence-specific concepts.

All these techniques have been developed primarily (or only) for the Java 
environment. It is, however, possible to interface them transparently to 
Frameworks built in other languages, like for example C++.

Fully functional prototypes of flexible and 
non-intrusive persistency modules have been build for several 
other packages, as for example FreeHEP AIDA and LCG Pool AttributeSet 
(package Indicium). 
\end{abstract}

\maketitle

\thispagestyle{fancy}

\section{JDO}

\subsection{Requirements on Transparent Persistence}

The Java Data Object (JDO)~\cite{JDO1},\cite{JDO2},\cite{Standard},\cite{Portal} 
standard has been created to satisfy several requirements on
the object persistence in Java:
\begin{itemize}
\item {\bf Object Model independence on persistency}:
  \begin{itemize}
  \item Java types are automatically mapped to native storage types.
  \item 3rd party objects can be persistified (even when their source is not
    available).
  \item The source of the persistent class is the same as the source of the transient
    class. No additional code is needed to make a class persistent.
  \item All classes can be made persistent (if it has a sense).
  \end{itemize}
\item {\bf Illusion of in-memory access to data}:
  \begin{itemize}
  \item Dirty instances (i.e. objects which have been changed after they have
    been read) are implicitly updated in the database.
  \item Catching, synchronization, retrieval and lazy loading are done 
    automatically.
  \item All objects, referenced from a persistent object, are automatically
    persistent ({\em Persistence by reachability}).
  \end{itemize}
\item {\bf Portability across technologies}:
  \begin{itemize}
  \item A wide range of storage technologies (relational databases, object-oriented
    databases, files,\dots) can be transparently used.
  \item All JDO implementations are exchangeable.
  \end{itemize}
\item {\bf Portability across platforms} is automatically available in
  Java.
\item {\bf No need for a different language} (DDL, SQL,\dots) to handle
  persistency (incl. queries).
\item {\bf Interoperability with Application Servers} (EJB~\cite{EJB},\dots).
\end{itemize}

\subsection{Architecture of Java Data Objects}

The Java Data Objects standard (Java Community Process Open Standard JSR-12)~\cite{Standard} has
been created to satisfy the requirements listed in the previous paragraph.

\begin{figure*}[!]
\centering
\includegraphics[width=135mm]{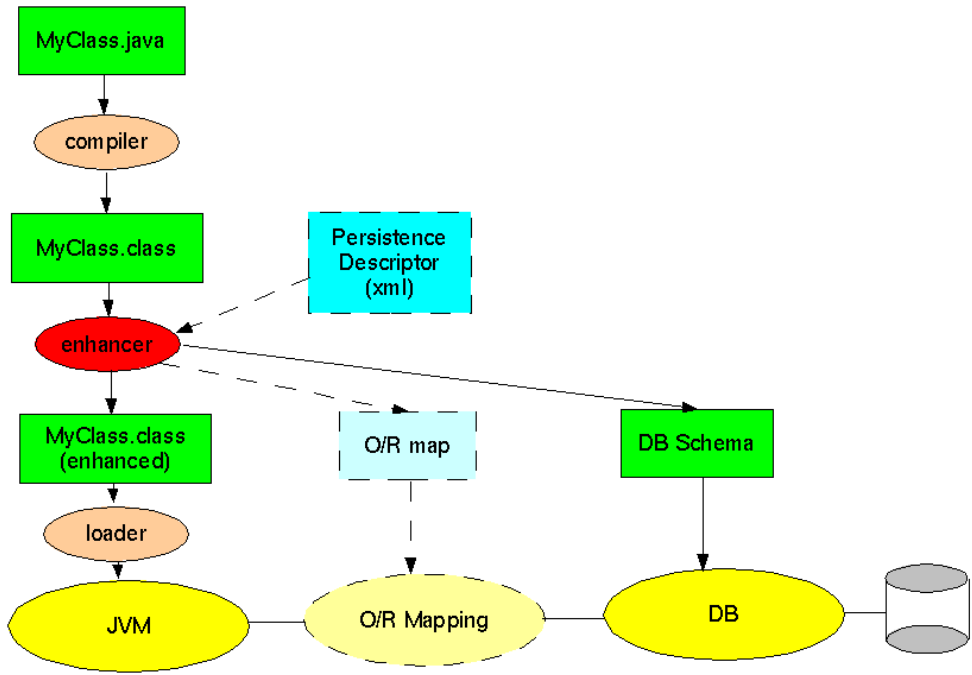}
\caption{JDO Enhancement.} 
\label{Enhancement}
\end{figure*}

\begin{figure}[!]
\centering
\includegraphics[width=80mm]{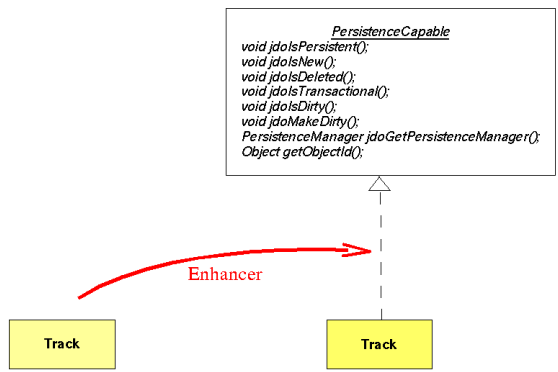}
\caption{Enhancer makes any class PersistenceCapable.} 
\label{PersistenceCapable}
\end{figure}

The persistence capability is added to a class by the Enhancer 
(as shown in Figure~\ref{Enhancement}):
\begin{itemize}
\item Enhancer makes a transient class PersistenceCapable by adding it all data and
  methods needed to provide the persistence functionality. After enhancement,
  the class implements PersistenceCapable interface (as shown in Figure~\ref{PersistenceCapable}).
\item Enhancer is generally applied to a class-file, but it can be also part
  of a compiler or a loader.
\item Enhancing effects can be modified via Persistence Descriptor (XML file). 
\item All enhancers are compatible. Classes enhanced with one JDO 
  implementation will work automatically with all other implementations.
\end{itemize}

\begin{figure}[!]
\centering
\includegraphics[width=80mm]{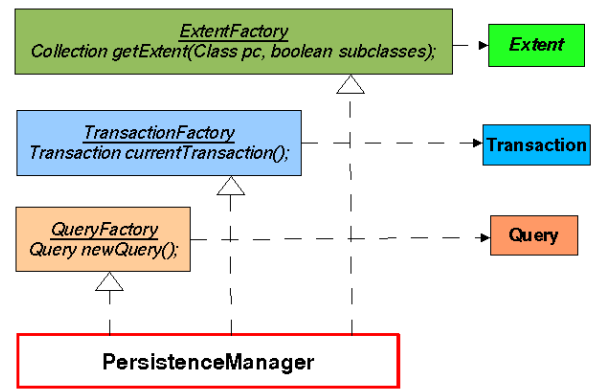}
\caption{All interactions with JDO are mediated by PersistenceManager.} 
\label{PersistenceManager}
\end{figure}

The main object, a user interacts with, is the PersistenceManager. It mediates 
all interactions with the database,
it manages instances lifecycle and it serves as a factory for Transactions,
Queries and Extents (as described in Figure~\ref{PersistenceManager}).

\subsection{Available Implementations}

After about a year since the JDO standardization, there are already many 
implementations available supporting all existing storage technologies.

\subsection{JDO Implementations}

\subsubsection{Commercial JDO Implementations}

Following commercial implementations of JDO standard exist:

enJin(Versant), FastObjects(Poet), FrontierSuit(ObjectFrontier), 
IntelliBO (Signsoft), JDOGenie(Hemisphere), JRelay(Object Industries), 
KODO(SolarMetric), LiDO(LIBeLIS), OpenFusion(Prism), Orient(Orient), 
PE:J(HYWY), \dots

These implementation often have a free community license available.

\subsubsection{Open JDO Implementations}

There are already several open JDO implementations available:
\begin{itemize}
\item {\bf JDORI}~\cite{JDORI} (Sun) is the reference and standard implementation. It 
  currently only works with the FOStore files. Support for a relational database
  via JDBC implementation is under
  development. It is the most standard, but not the most performant implementation.
\item {\bf TJDO}~\cite{TJDO} (SourceForge) is a high quality implementation originally written
  by the TreeActive company, later put on the GPL license. It supports all important
  relational databases. It supports an automatic creation of the database schema. It
  implements full JDO standard.
\item {\bf XORM}~\cite{XORM} (SourceForge) does not yet support full JDO standard. It 
  does not automatically generate a database schema, on the other hand, it allows
  a reuse of existing schemas.
\item {\bf JORM}~\cite{JORM} (JOnAS/ObjectWeb) has a fully functional object-relational
  mapping, the full JDO implementation is under development.
\item {\bf OJB}~\cite{OJB} (Apache) has a mature object-relational engine. Full JDO interface
  is not yet provided.
\end{itemize}

\subsection{Supported Databases}

All widely used databases are already supported either by their provider or by
a third party:
\begin{itemize}
\item {\bf RDBS and ODBS}: Oracle, MS SQL Server, DB2, PointBase, Cloudscape,
  MS Access, JDBC/ODBC Bridge, Sybase, Interbase, InstantDB, Informix, SAPDB,
  Postgress, MySQL, Hypersonic SQL, Versant,\dots
\item {\bf Files}: XML, FOSTORE, flat, C-ISAM,\dots
\end{itemize}
The performance of JDO implementations is determined by the native performance of
a database. JDO itself introduces a very small overhead.

\section{HEP Applications using JDO}

\subsection{Trivial Application}

A simple application using JDO to write and read data is shown in Listing~\ref{Trivial}.

\begin{table*}[!]
\begin{center}
\begin{tabular}{|l|}
\hline
$//\ Initialization$ \\
$PersistenceManagerFactory\ pmf = JDOHelper.getPersistenceManagerFactory(properties);$ \\
$PersistenceManager\ pm = pmf.getPersistenceManager();$ \\
$Transaction\ tx = pm.currentTransaction();$ \\
\\
$//\ Writing$ \\
$tx.begin();$ \\
$\dots$ \\
$Event\ event = \dots;$ \\
$pm.makePersistent(event);$ \\
$\dots$ \\
$tx.commit();$ \\
\\
$//\ Searching\ using\ Java-like\ query\ language\ translated\ internally\ to\ DB\ native\ query\ language$ \\
$//\ (SQL\ available\ too\ for\ RDBS)$ \\
$tx.begin();$ \\
$Extent\ extent = pm.getExtent(Track.class, true);$ \\
$String\ filter = "pt > 20.0";$ \\
$Query\ query = pm.newQuery(extent, filter);$ \\
$Collection\ results = query.execute();$ \\
$\dots$ \\
$tx.commit();$ \\
\hline
\end{tabular}
\caption{Trivial example of using JDO.}
\label{Trivial}
\end{center}
\end{table*}

\subsection{Indicium}

Indicium~\cite{Indicium} has been created to satisfy the LCG~\cite{LCG} 
Pool~\cite{Pool} requirements on the Metadata management:
``To define, accumulate, search, filter and manage Attributes (Metadata)
external/additional to existing (Event) data.'' Those metadata are a generalization
of the traditional Paw ntuple concept. They are used in the first phase of the analysis
process to make a pre-selection of Event for further processing. They should be
efficient. They are apparently closely related to Collections (of Events).

The Indicium package provides an implementation of the AttributeSet (Event 
Metadata, Tags) for the LCG/Pool project in Java and C++ (with the same API). 
The core of Indicium is implemented in Java.

All expressed requirements can only be
well satisfied by the system which allows in principle any object to act
as an AttributeSet. Such system can be easily built when we realize that 
mentioned requirements are satisfied by JDO:
\begin{itemize}
\item {\bf AttributeSet} is simply any Object with a reference to another
  (Event) Object.
\item {\bf Explicit Collection} is just any standard Java Collection.
\item {\bf Implicit Collection} (i.e. all objects of some type T within a
  Database) is directly the JDO Extent.
\end{itemize}

Indicium works with any JDO/DB implementation. As all the requirements are
directly satisfied by the JDO itself, the Indicium only implements a simple
wrapper and a code for database management (database creation, opening, \dots).
That is in fact the only database-specific code.

It is easy to switch between various JDO/DB implementations via a simple
properties file. The default Indicium implementation contains configuration for
JDORI with FOStore file format and TJDO with Cloudscape or MySQL
databases, others are simple to add.

The data stored by Indicium are accessible also via native database protocols
(like JDBC or SQL) and tools using them.

As it has been already mentioned, Indicium provides just a simple convenience
layer on top of JDO trying to capture standard AttributeSet usage 
patterns. There are four ways how AttributeSet can be defined:
\begin{itemize}
\item {\bf Assembled} AttributeSet is fully constructed at run-time in a way
  similar to classical Paw ntuples.
\item {\bf Generated} AttributeSet class is generated from a simple XML
  specification.
\item {\bf Implementing} AttributeSet can be written by hand to implement
  the standard AttributeSet Interface.
\item {\bf FreeStyle} AttributeSet can be just about any class. It can be managed
  by the Indicium infrastructure, only some convenience functionality may be
  lost.
\end{itemize}

\begin{figure*}[!]
\centering
\includegraphics[width=135mm]{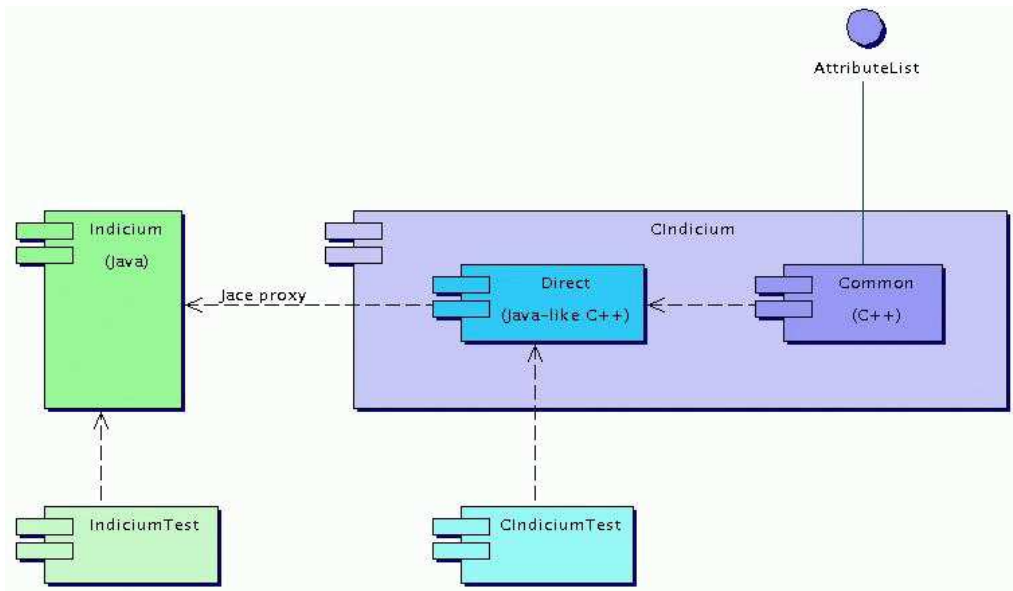}
\caption{CIndicium - C++ interface to Indicium.} 
\label{AttributeSet}
\end{figure*}

To satisfy also the requirements of C++ users, the C++ interface of Indicium has been
created in the form of JACE~\cite{JACE} proxies. This way, C++ users can directly use Indicium
Java classes from a C++ program. CIndicium Architecture is shown in Figure~\ref{AttributeSet},
an example of its use is shown in Listing~\ref{CIndicium}.

\begin{table*}[!]
\begin{center}
\begin{tabular}{|l|}
\hline
$//\ Construct\ Signature$ \\
$Signature\ signature("AssembledClass");$ \\
$signature.add("j", "int", "Some Integer Number");$ \\
$signature.add("y", "double", "Some Double Number");$ \\
$signature.add("s", "String", "Some String");$ \\
\\
$//\ Obtain\ Accessor\ to\ database$ \\
$Accessor\ accessor = AccessorFactory::createAccessor("MyDB.properties");$ \\
\\
$//\ Create\ Collection$ \\
$accessor.createCollection("MyCollection", signature, true);$ \\
\\
$//\ Write\ AttributeSets\ into\ database$ \\
$AssembledAttributeSet*\ as;$ \\
$for (int\ i = 0; i < 100; i++) \{$ \\
$\ \ as = new\ AssembledAttributeSet(signature);$ \\
$\ \ as->set("j", ...);$ \\
$\ \ as->set("y", ...);$ \\
$\ \ as->set("s", ...);$ \\
$\ \ accessor.write(*as);$ \\
$\ \ \}$ \\
\\
$//\ Search\ database$ \\
$std::string\ filter = "y > 0.5";$ \\
$Query\ query = accessor.newQuery(filter);$ \\
$Collection\ collection  = query.execute();$ \\
$std::cout << "First: " << collection.toArray()[0].toString() << std::endl;$ \\
\hline
\end{tabular}
\caption{Example of CIndicium use.}
\label{CIndicium}
\end{center}
\end{table*}

\subsection{AIDA Persistence}

JDO has been used to provide a basic persistency service for the FreeHEP~\cite{FreeHEP} reference
implementation of AIDA~\cite{AIDA}. Three kinds of extension to the existing implementation
have been required:
\begin{itemize}
\item Implementation of the IStore interface as AidaJDOStore.
\item Creation of the XML description for each AIDA class (for example see  Listing~\ref{AIDA}).
\item Several small changes to exiting classes, like creation of wrappers around
  arrays of primitive types, etc.
\end{itemize}

\begin{table*}[!]
\begin{center}
\begin{tabular}{|l|}
\hline
$<jdo>$ \\
$\ \ <package\ name="hep.aida.ref.histogram">$ \\
$\ \ \ \ <class\ name="Histogram2D"$ \\
$\ \ \ \ \ \ \ persistence-capable-superclass="hep.aida.ref.histogram.Histogram">$ \\
$\ \ \ \ \ \ \ </class>$ \\
$\ \ \ \ </package>$ \\
$\ \ </jdo>$ \\
\hline
\end{tabular}
\caption{Example of JDO descriptor for AIDA class.}
\label{AIDA}
\end{center}
\end{table*}

It has become clear, that the AIDA persistence API is not sufficient and it has to
be made richer to allow more control over persistent objects, better searching
capabilities, etc.

\subsection{Minerva}

Minerva~\cite{Minerva} is a lightweight Java Framework which implements main Architectural
principles of the ATLAS C++ Framework Athena~\cite{Athena}:
\begin{itemize}
\item {\bf Algorithm - Data Separation}: Algorithmic code is separated from 
  the data on which it operates. Algorithms can be explicitly called and 
  don't a have persistent state (except for parameters). Data are potentially 
  persistent and processed by Algorithms.
\item {\bf Persistent - Transient Separation}: The Persistency mechanism is 
  implemented by specified components and have no impact on the 
  definition of the transient Interfaces. Low-level Persistence technologies 
  can be replaced without changing the other Framework components (except 
  for possible configuration). A specific definition of Transient-Persistent 
  mapping is possible, but is not required.
\item {\bf Implementation Independence}: There are no implementation-specific 
  constructs in the definition of the interfaces. In particular, all 
  Interfaces are defined in an implementation independent way. Also all 
  public objects (i.e. all objects which are exchanged between components 
  and which subsequently appear in the Interface' definitions) are 
  identifiable by implementation independent Identifiers.
\item {\bf Modularity}: All components are explicitly designed with 
  interchangeability in mind. This implies that the main deliverables are 
  simple and precisely defined general interfaces and existing implementation 
  of various modules serves mainly as a Reference implementation.
\end{itemize}

\begin{figure*}[!]
\centering
\includegraphics[width=135mm]{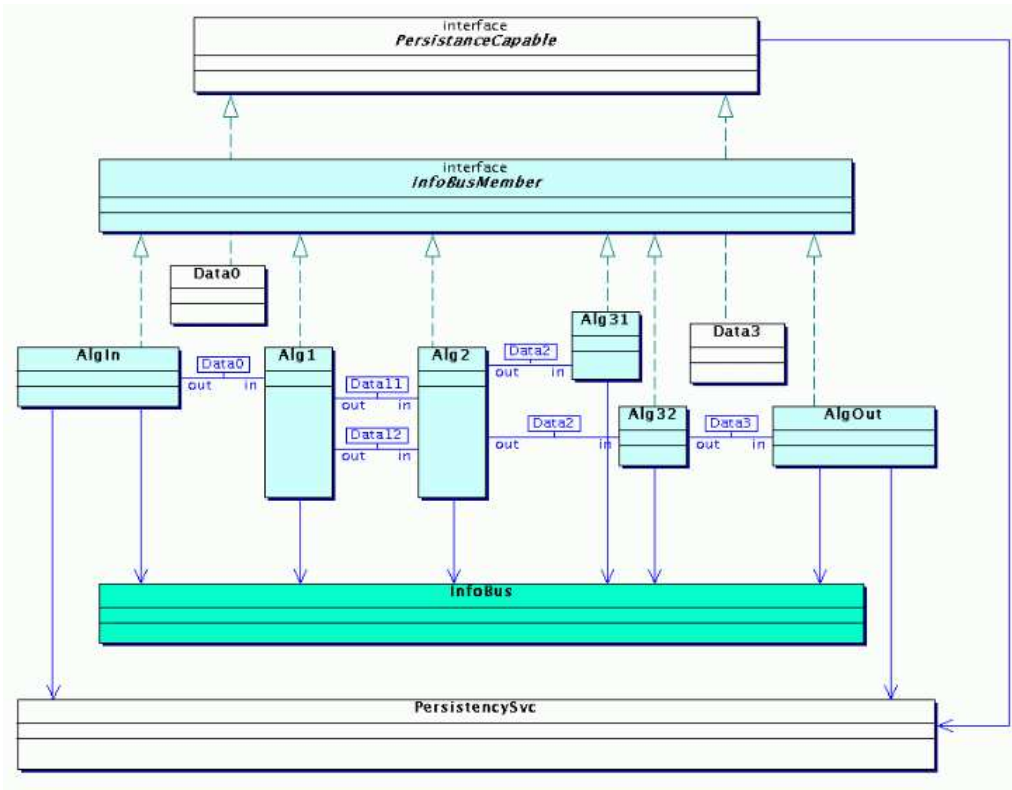}
\caption{Minerva is based on the InfoBus scheduling and the JDO persistency.} 
\label{InfoBus}
\end{figure*}

Minerva scheduling is based on InfoBus~\ref{InfoBus}  Architecture:
\begin{itemize}
\item Algorithms are {\em Data Producers} or {\em Data Consumers} (or both).
\item Algorithm declare their supported I/O types.
\item Scheduling is done implicitly. An Algorithm runs when it has all its
  inputs ready.
\item Both Algorithms and Services run as (static or dynamic) Servers.
\item The environment is naturally multi-threaded.
\end{itemize}

Overview of the Minerva Architecture is shown in Figure~\ref{InfoBus}.

\begin{figure}[!]
\centering
\includegraphics[width=80mm]{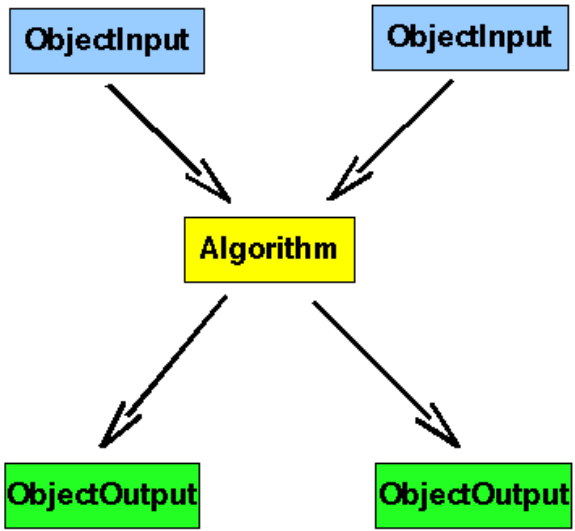}
\caption{Example of a Minerva run.} 
\label{Minerva}
\end{figure}

It is very easy to configure and run Minerva. For example, one can create
a Minerva run with 5 parallel Servers. Two of them are reading Events from
two independent databases, one is processing each Event and two last write
new processed Events on two new databases depending on the Event characteristics.
(See Figure~\ref{Minerva} for a schema of such run and Listing~\ref{MinervaScript} 
for its steering script.)

\begin{table}[!]
\begin{center}
\begin{tabular}{|l|}
\hline
$new\ Algorithm(<Algorithm\ properties>);$ \\
$new\ ObjectOutput(<db3>, <Event\ properties1>);$ \\
$new\ ObjectOutput(<db4>, <Event\ properties2>);$ \\
$new\ ObjectInput(<db1>);$ \\
$new\ ObjectInput(<db2>);$ \\
\hline
\end{tabular}
\caption{Example of steering script for a Minerva run.}
\label{MinervaScript}
\end{center}
\end{table}

\begin{figure*}[!]
\centering
\includegraphics[width=135mm]{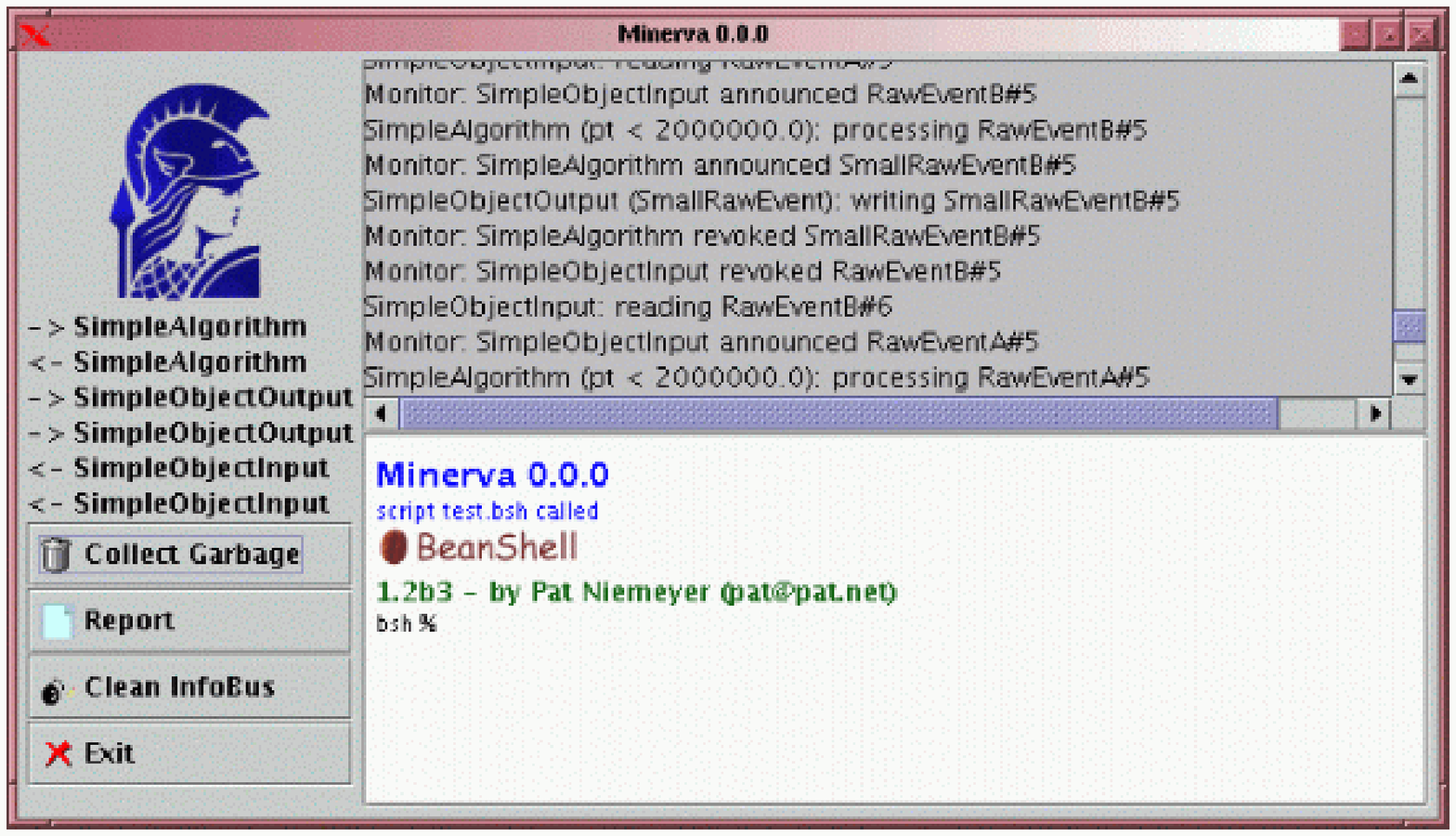}
\caption{Running set of Producers and Consumers created from a script inside
  Minerva.} 
\label{GUI}
\end{figure*}

\begin{figure*}[!]
\centering
\includegraphics[width=135mm]{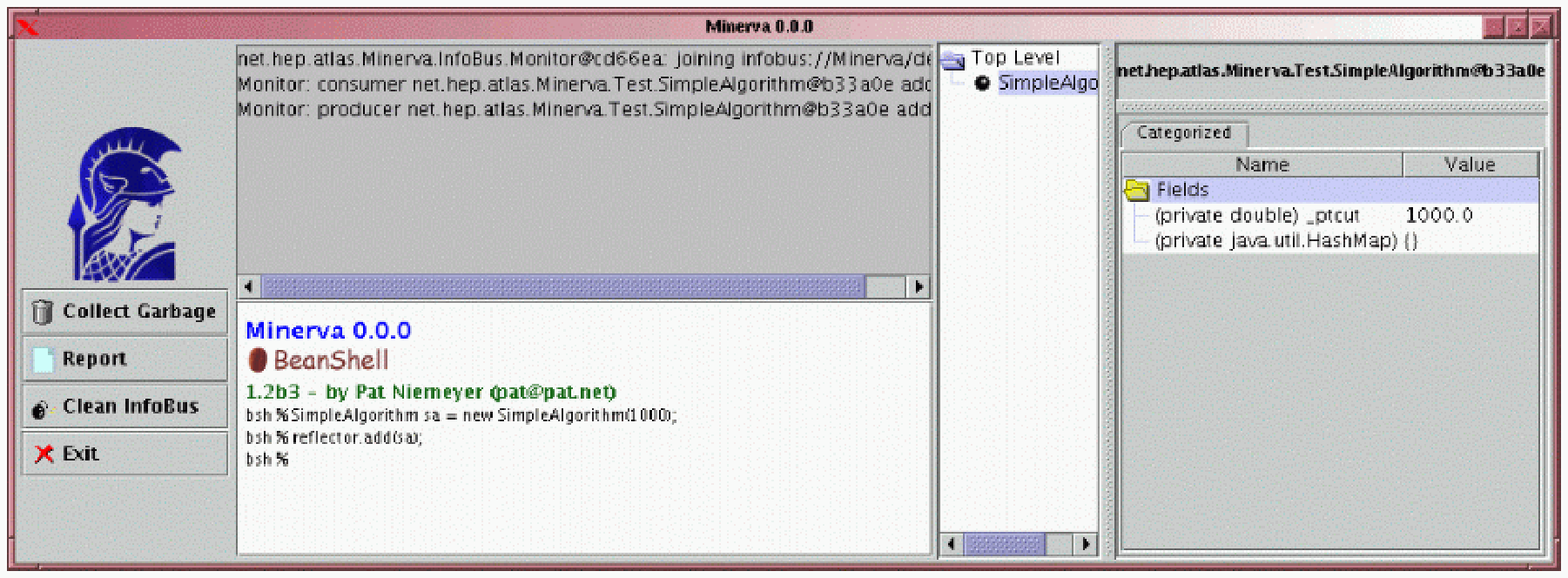}
\caption{Using ObjectBrowser to inspect Algorithm inside Minerva.} 
\label{ObjectBrowser}
\end{figure*}

Minerva has also simple but powerful modular Graphical User Interface which 
allows to plug in easily other components as the BeanShell~\cite{BeanShell} command-line interface,
the JAS~\cite{JAS} histogramming, the ObjectBrowser~\cite{ObjectBrowser}, etc. 
Figure~\ref{GUI} and Figure~\ref{ObjectBrowser} show examples of running Minerva with various 
interactive plugins loaded.

\section{Prototypes using JDO}

\subsection{Object Evolution}

It is often necessary to change object' shape while keeping its content and
identity. This functionality is especially needed in the persistency
domain to satisfy {\em Schema Evolution} (Versioning) or {\em Object Mapping}
(DB Projection), i.e. retrieving an Object of type A dressed as an Object of
another type B. This functionality is not addressed by JDO. In practice, it is
handled either on the lower lever (in a database) or on the higher level (in
the overall framework, for example EJB).

It is, however, possible to implement an Object Evolution for JDO with the
help of Dynamic Proxies and Aspects.

\begin{figure}[!]
\centering
\includegraphics[width=80mm]{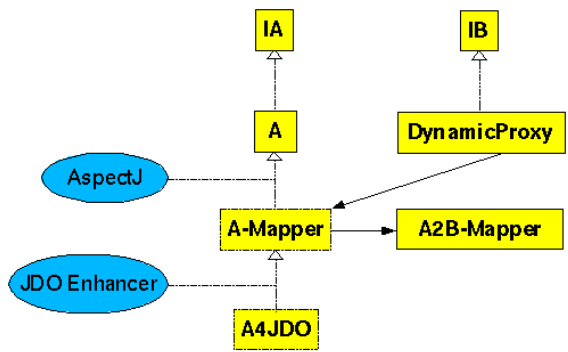}
\caption{Support for Object Evolution.} 
\label{Evolution}
\end{figure}

Let's suppose that a user wants to read an Object of a type A (of an 
Interface IA) dressed as an Object of another Interface IB. To enable that,
four components should co-operate (as shown in Fig~\ref{Evolution}):
\begin{itemize}
\item JDO Enhancer enhances class A so it is PersistenceCapable and it is
  managed by JDO PersistenceManager.
\item AspectJ~\cite{AspectJ} adds read-callback with the mapping A $\rightarrow$ IB. This is called
  automatically when JDO reads an object A.
\item A simple database of mappers provides a suitable mapping between A 
  and IB.
\item DynamicProxy delivers the content of the Object A with the interfaces IB:\\
  $IB\ b = (IB)DynamicProxy.newInstance(A, IB);$.
\end{itemize}
All those manipulations are of course hidden from the End User.

\subsection{Foreign References}

HEP data are often stored in sets of independent databases, each one managed
independently. This architectures do not directly support references between
objects from different databases (while references inside one database are
managed directly by the JDO support for Persistence by Reachability). As in the case
of the Object Evolution, foreign references are usually resolved either on
the lower level (i.e. all databases are managed by one storage manager and
JDO operates on top) or on the higher level (for example by the EJB framework).

Another possibility is to use a similar Architecture as in the case of Object
Evolution with Dynamic Proxy delivering foreign Objects.

\begin{figure}[!]
\centering
\includegraphics[width=80mm]{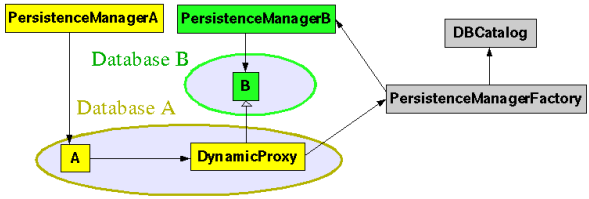}
\caption{Support for Foreign References.} 
\label{References}
\end{figure}

Let's suppose, that a User reads an object A, which contains a reference
to another object B, which is actually stored in a different database (and
thus managed by a different PersistenceManager). The database with the object A
doesn't in fact in this case contain an object B, but a DynamicProxy object.
The object B can be transparently retrieved using three co-operating components 
(as shown on Fig~\ref{References}):
\begin{itemize}
\item When reference from an object A to an object B is requested, JDO delivers 
  DynamicProxy instead.
\item The DynamicProxy asks PersistenceManagerFactory for a PersistenceManager
  which handles the object B. It then uses that PersistenceManager to get the 
  object B and casts itself into it.
\item PersistenceManagerFactory gives this information by interrogating
  DBcatalog (possibly a Grid Service).
\end{itemize}
All those manipulations are of course hidden from the End User.

\section{Summary}

It has been shown that JDO standard provides suitable foundation of the 
persistence service for HEP applications.

Two major characteristics of persistence solutions based on JDO are:
\begin{itemize}
\item Not intrusiveness.
\item Wide range of available JDO implementations, both commercial and free, 
  giving access to all major databases.
\end{itemize}
JDO profits from the native databases functionality and performance (SQL 
queries,...), but presents it to users in a native Java API.

\newpage
\onecolumngrid

\end{document}